\documentclass[journal]{IEEEtran}

\usepackage{cite,graphicx,amssymb,amsmath,color,textcomp,subfigure,multirow}

\begin{document}

\title{ Channel Estimation for Wireless Communication Systems Assisted by Large Intelligent Surfaces}

\author{Junliang Lin, Gongpu Wang, Rongfei Fan, Theodoros A. Tsiftsis, and Chintha Tellambura

\thanks{J. Lin and G. Wang are with Beijing Key Lab of Transportation Data Analysis
and Mining, School of Computer and Information Technology, Beijing
Jiaotong University, Beijing 100044, China (e-mail: 19112019,~gpwang@bjtu.edu.cn).}

\thanks{R. Fan is with the School of Information and Electronics, Beijing Institute of Technology, Beijing 100081, China (e-mail: fanrongfei@bit.edu.cn).}

\thanks{T. A. Tsiftsis is with the School of Intelligent Systems Science and Engineering, Jinan University, Zhuhai 519070, China (e-mail:
theodoros.tsiftsis@gmail.com).}

\thanks{C. Tellambura is with the Department of Electrical and Computer Engineering, University of Alberta, Edmonton, AB T6G 2V4, Canada (e-mail:
chintha@ece.ualberta.ca).}
}
\maketitle

\begin{abstract}
In this letter, the channel estimation problem is studied for wireless communication systems assisted by large intelligent surface. Due to features of assistant channel, channel estimation (CE) problem for the investigated system is shown as a constrained estimation error minimization problem, which differs from traditional CE problems. A Lagrange multiplier and dual ascent-based estimation scheme is then designed to obtain a closed-form solution for the estimator iteratively. Moreover, the Cram\'er-Rao lower bounds are deduced for performance evaluation. Simulation results show that the designed scheme could improve estimation accuracy up to $18\%$, compared with least square method in low signal-to-noise ratio regime.
\end{abstract}


\section{Introduction}
\IEEEPARstart{C}{urrently}, increasing energy consumption and hardware expenditures are two great challenges facing wireless communication systems. Radio environment assisted by large intelligent surface (LIS) is foreseen as a promising paradigm to tackle the issues \cite{Hu}. LIS is a passive meta-surface which is made up of multiple low-cost reflecting elements. These elements could be appropriately configured to adjust the phase of incident signals with low energy consumption according to the channel state information (CSI) of both base station (BS)-LIS links and LIS-user equipment (UE) links.

Accurate channel estimation (CE) is considered as one of the most essential issues in wireless communications. CE for the emerging wireless communication systems assisted by LIS is recently drawing attention of academia. An optimal binary reflection controlled CE protocol was designed in \cite{Mishra} for LIS-assisted system to maximize the practical efficacy. In \cite{Nadeem}, a minimum mean squared error-based CE protocol was presented by employing a time division structure. CE problem was analysed in \cite{Taha} utilizing compressive sensing and deep learning.

Prior works focus on estimating either each channel parameter sequentially \cite{Mishra,Nadeem}, or performing accurate channel estimation with the help of large pre-collected training data \cite{Taha}. Besides, we notice that the impact of the features of assistant channels on CE has not been investigated yet, which motivates this work.

In this letter, we address CE for wireless communication systems assisted by LIS. We formulate a CE problem as a constrained optimization problem considering the features of assistant channels, which is different from the traditional CE problems. Then, an effective estimation scheme is designed for acquiring the closed-form expression for the channel parameters in an iterative manner. Furthermore, we deduce the corresponding Cram\'er-Rao lower bound (CRLB) and compare with the least square (LS) method through numerical results.


\section{System Model and Channel Model}

\begin{figure}[t]
\centering
\includegraphics[scale=0.7]{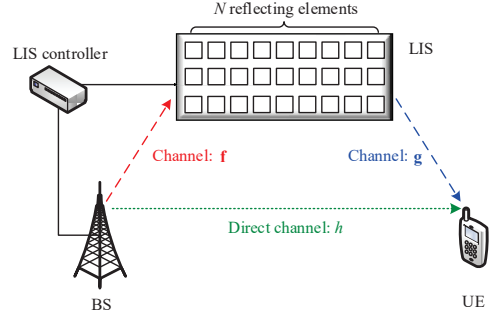}
\caption{A wireless system assisted by LIS where the signal from BS could be transmitted through direct channel $h$ (green dot line) and/or channels $\bf f$ (red dash line) and $\bf g$ (blue dash line). }
\label{fig:System_Model}
\end{figure}

Consider a wireless system in Fig. \ref{fig:System_Model}, which consists of a single-antenna BS, a single-antenna UE, a controller and a LIS. The LIS has $N$ reflecting elements, and is set up to assist the downlink transmission from BS to UE. The controller is responsible for tuning the phases of these reflecting elements through the interaction with BS.

Denote the direct channel from BS to UE, the channels from BS to LIS, the channels from LIS to UE by $h$, ${\bf f}\in {\mathbb C}^{N\times1}$ and ${\bf g}\in {\mathbb C}^{N\times1}$, whose elements follow a complex zero-mean Gaussian distribution with variances $\sigma_h^2$, $\sigma_f^2$ and $\sigma_g^2$, respectively. Let $r[k]\in\{0,1\}$ be the indicator identifying whether LIS reflects the $k$th symbol to UE. Specifically, if a symbol is not reflected by LIS, then $r[k]=0$; otherwise, $r[k]=1$. In the case of $r[k]=0$, the received symbol takes the structure
\begin{equation}\label{eq:yrk0}
y[k] = hs[k] + w[k]
\end{equation}
where $s[k]$ denotes the $k$th transmitted symbol, and $w[k]$ models the complex zero-mean Gaussian noise at UE with variance $\sigma_w^2$. When $r[k]=1$, the received symbol equals
\begin{equation}\label{eq:yrk1}
y[k] = h s[k] + {\bf f}^{\rm T} {\bf \Phi} {\bf g} s[k] + w[k]
\end{equation}
where ${\boldsymbol \Phi}=\text{diag}(e^{j\phi_1},e^{j\phi_2},\dots,e^{j\phi_N})$ is the phase-shift matrix with $\phi_n$ characterizing the tunable phase induced by the $n$th reflecting element. For simplicity, we define the assistant channel as
\begin{equation}\label{eq:eta}
\eta \buildrel \Delta \over = {\bar{\bf f}}^{\rm T}{\bf \Theta}{\bf \Phi}{\bf \Psi}{\bar{\bf g}} \buildrel (a) \over ={\bar{\bf f}}^{\rm T} {\bar{\bf g}}
\end{equation}
where $\bar{\bf{f}} = {(|{f_1}|,|{f_2}|, \ldots ,|{f_N}|)^{\rm T}}, \bar{\bf{g}} = {(|{g_1}|,|{g_2}|, \ldots ,|{g_N}|)^{\rm T}}$ denote the amplitude vector of fading channels $\bf f$ and $\bf g$, respectively; ${\boldsymbol \Theta}=\text{diag}(e^{-j\theta_1},e^{-j\theta_2},\dots,e^{-j\theta_N})$ and ${\boldsymbol \Psi}=\text{diag}(e^{-j\psi_1},e^{-j\psi_2},\dots,e^{-j\psi_N})$ are phase matrices with $\theta_n$ and $\psi_n$ depicting the phase of the $n$th element in $\bf f$ and $\bf g$, respectively. In addition, step $(a)$ assumes that $\bf \Phi$ could be configured to eliminate $\boldsymbol \Theta$ and $\boldsymbol \Psi$ with perfect CSI, i.e., ${\bf \Theta}{\bf \Phi}{\bf \Psi}=\bf I$ \cite{Basar}. Clearly, we have $\eta>0$.

Combining (\ref{eq:yrk0}), (\ref{eq:yrk1}) with (\ref{eq:eta}), we express the received symbol as
\begin{equation}\label{eq:yk}
y[k]= \left\{ {\begin{array}{*{20}{l}}
{hs[k] + w[k],}&{s[k]\in{\rm pilot}, r[k]=0}\\
{hs[k] + {\eta} s[k] + w[k],}&{s[k]\in{\rm pilot}, r[k]=1}\\
{hs[k] + r[k]{\eta} s[k] + w[k],}&{s[k]\in{\rm data}}
\end{array}} \right..
\end{equation}

In this letter, we aim to estimate the channel parameters $h$ and $\eta$, and investigate the impact of features of assistant channel on CE performance by considering (i) the nature of assistant channel, i.e., $\eta>0$, and (ii) the superiority of assistant channel over direct channel with sufficient reflecting elements \cite{Zhao2}, i.e., $\eta>|h|$.

\section{Channel Estimation}

In this section, we design a Lagrange multiplier and dual ascent-based estimation scheme to address the constrained CE problem. Then, we derive the corresponding CRLB for performance evaluation.

\subsection{Scheme Design}
\begin{figure}[t]
\centering
\includegraphics[scale=0.7]{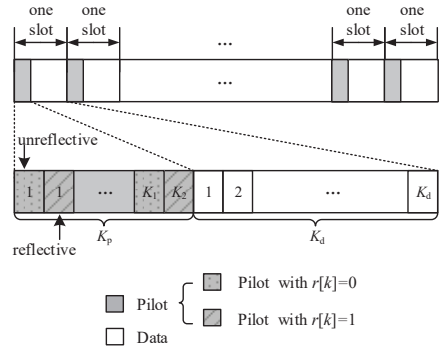}
\caption{Slotted structure where grey dot blocks denote pilot with $r[k]=0$, grey blocks with oblique line represent the pilot with $r[k]=1$ and white blocks stand for data symbol, respectively. }
\label{fig:Pilot Structure}
\end{figure}

Suppose the transmission follows a slotted structure, as depicted in Fig. \ref{fig:Pilot Structure}. Each slot contains $K_{\rm p}$ pilot symbols and $K_{\rm d}$ data symbols. Within $K_{\rm p}$ pilots, the number of symbols that LIS remains unreflective and reflective are $K_1$ and $K_2$, respectively.

Specifying two index sets ${\mathcal K}_1=\{u_1,u_2,\dots,u_{K_1}\}$ and ${\mathcal K}_2=\{v_1,v_2,\dots,v_{K_2}\}$, we define pilots with $r[k]=0$ and $r[k]=1$ as
\begin{equation}\label{eq:s1s2}
\begin{split}
&{{\bf{s}}_{\rm p1}} \buildrel \Delta \over = {\left[ {s\left[ {{u_1}} \right],s\left[ {{u_2}} \right], \ldots ,s\left[ {{u_{{K_1}}}} \right]} \right]^{\rm T}},\\
&{{\bf{s}}_{\rm p2}} \buildrel \Delta \over = {\left[ {s\left[ {{v_1}} \right],s\left[ {{v_2}} \right], \ldots ,s\left[ {{v_{{K_2}}}} \right]} \right]^{\rm T}},\\
\end{split}
\end{equation}
with corresponding noises and received signals given by
\begin{equation}\label{eq:w1y1w2y2}
\begin{split}
&{{\bf{w}}_{\rm p1}} \buildrel \Delta \over = {\left[ {w\left[ {{u_1}} \right],w\left[ {{u_2}} \right], \ldots ,w\left[ {{u_{{K_1}}}} \right]} \right]^{\rm T}},\\
&{{\bf{y}}_{\rm p1}} \buildrel \Delta \over = {\left[ {y\left[ {{u_1}} \right],y\left[ {{u_2}} \right], \ldots ,y\left[ {{u_{{K_1}}}} \right]} \right]^{\rm T}},\\
&{{\bf{w}}_{\rm p2}} \buildrel \Delta \over = {\left[ {w\left[ {{v_1}} \right],w\left[ {{v_2}} \right], \ldots ,w\left[ {{v_{{K_2}}}} \right]} \right]^{\rm T}},\\
&{{\bf{y}}_{\rm p2}} \buildrel \Delta \over = {\left[ {y\left[ {{v_1}} \right],y\left[ {{v_2}} \right], \ldots ,y\left[ {{v_{{K_2}}}} \right]} \right]^{\rm T}}.
\end{split}
\end{equation}

With the aid of (\ref{eq:s1s2}) and (\ref{eq:w1y1w2y2}), the pilot-based received signal in (\ref{eq:yk}) is rewritten in vector form:
\begin{equation}\label{eq:y1y2}
\begin{split}
&{\bf{y}_{\rm p1}} = h{\bf s}_{\rm p1} + {\bf w}_{\rm p1},\\
&{\bf{y}_{\rm p2}} = h{\bf s}_{\rm p2}+ \eta{\bf s}_{\rm p2} + {\bf w}_{\rm p2}.
\end{split}
\end{equation}

Consequently, we reconstruct the received signal based on (\ref{eq:y1y2}) as
\begin{equation}\label{eq:bfy}
{\bf{y}} = {\bf A}{\bf x} + {\bf w}
\end{equation}
where ${\bf x}=[h,\eta]^{\rm T}\in {\mathbb C}^{2\times1}$, ${\bf A} = \left[ {\begin{array}{*{20}{c}}
{{{\bf{s}}_{\rm p1}}}&{{{{\bf0}_{{K_1}\times1}}}}\\
{{{\bf{s}}_{\rm p2}}}&{{{\bf{s}}_{\rm p2}}}
\end{array}} \right]\in {\mathbb C}^{K_{\rm p}\times2}$, ${\bf w}= {[{\bf{w}}_{\rm p1}^{\rm T},{\bf{w}}_{\rm p2}^{\rm T}]^{\rm T}}\in {\mathbb C}^{K_{\rm p}\times1}$ and ${\bf y}= {[{\bf{y}}_{\rm p1}^{\rm T},{\bf{y}}_{\rm p2}^{\rm T}]^{\rm T}}\in {\mathbb C}^{K_{\rm p}\times1}$ denote the channel parameter vector, pilot matrix, noise vector and pilot-based received signal vector, respectively.

With the signal model in (\ref{eq:bfy}), we formulate the CE problem as a constrained estimation error minimization problem by considering the features of assistant channel, i.e.,
\begin{equation}\label{eq:P}
\begin{split}
\mathop {\min}\limits_{{\bf x}}~~&||{\bf{y}}- {\bf A}{\bf x}||^2\\
\mathrm{s.t.}~~~ &{\rm C1}:{\eta}>0, \\
~~~&{\rm C2}:\eta>| h|.
\end{split}
\end{equation}

As optimization problem (\ref{eq:P}) is thorny to address due to the existence of ${\rm C2}$, we square both sides of the inequality, i.e.,
\begin{equation}
{\rm C2}:\eta^2>|h|^2.
\end{equation}

Denote ${\bf b}=[0,-1]^{\rm T}$ and ${\bf C}=\left[ {\begin{array}{*{20}{c}}
{1}&{0}\\
{0}&{-1}
\end{array}} \right]$, we further relax $\rm C1$ and $\rm C2$ to
\begin{equation}
{\rm C1}':{\bf b}^{\rm T}{\bf x}^*\le0,
\end{equation}
\begin{equation}
{\rm C2}':{\bf x}^{\rm H}{\bf C}{\bf x} \le 0.
\end{equation}

Therefore, the original optimization problem in (\ref{eq:P}) becomes
\begin{equation}\label{eq:P'}
\begin{split}
\mathop {\min}\limits_{{\bf x}}~~&||{\bf{y}}- {\bf A}{\bf x}||^2\\
\mathrm{s.t.}~~~ &{\rm C1}':{\bf b}^{\rm T}{\bf x}^*\le0, \\
~~~&{\rm C2}':{\bf x}^{\rm H}{\bf C}{\bf x} \le 0.
\end{split}
\end{equation}

Since the transformed optimization problem in (\ref{eq:P'}) is convex, we design a Lagrange multiplier and dual ascent-based method \cite{Boyd} to find the globally optimal estimator. To do so, we express the Lagrange function as
\begin{align}
L\left( {{\bf x}, {\bf x}^*, \lambda, \delta} \right) =& {{\bf y}^{\rm H}}{\bf y} - {{\bf y}^{\rm H}}{\bf A}{\bf x} - {{\bf x}^{\rm H}}{{\bf A}^{\rm H}}{\bf y} + {{\bf x}^{\rm H}}{{\bf A}^{\rm H}}{\bf A}{\bf x}\nonumber\\
& + \lambda{\bf b}^{\rm T}{\bf x}^* + \delta{\bf x}^{\rm H}{\bf C}{\bf x}
\end{align}
where $\lambda\ge0$ and $\delta\ge0$ are the Lagrange multipliers associated to $\rm C1'$ and $\rm C2'$, respectively. Therefore, the dual maximization problem is formulated as
\begin{equation}\label{eq:dualL}
\begin{split}
&\mathop {\max }\limits_{\lambda,\delta } ~\inf\limits_{{\bf x}} L({\bf x},{\bf x}^*,\lambda,\delta)\\
&~~{\rm s.t.} ~~~~~~~\lambda, \delta \ge 0.
\end{split}
\end{equation}

After differentiating $L(\bf x,\bf x^*,\lambda,\delta)$ with respect to $\bf x^*$, we obtain the necessary conditions for the optimal solution using the Karush-Kuhn-Tucker (KKT) conditions \cite{Boyd}. Thus, the optimal infimum of $\hat{\bf x}$ in closed-form expression can be derived as
\begin{equation}
{\hat{\bf x}}^{\star} = ({\bf A}^{\rm H}{\bf A}+{\bf C}\delta^{\star})^{-1}({\bf A}^{\rm H} {\bf y}-{\bf b}\lambda^{\star}).
\end{equation}
Also, by considering the dual ascent method being applied to solve (\ref{eq:dualL}), the following steps are conducted sequentially until convergence, i.e.,
\begin{equation}
\begin{split}
&{\hat{\bf x}}^{(t+1)} = ({\bf A}^{\rm H}{\bf A}+{\bf C}\delta^{(t)})^{-1}({\bf A}^{\rm H} {\bf y}-{\bf b}\lambda^{(t)}),\\
&\lambda^{(t+1)} = \left[\lambda^{(t)} + \varepsilon^{(t)} {\bf b}^{\rm T}{\hat{\bf x}^{*(t+1)}}\right]^+,\\
&\delta^{(t+1)} = \left[\delta ^{(t)} + \tau^{(t)}{\hat{\bf x}}^{{\rm H}(t+1)}{\bf C}{\hat{\bf x}}^{(t+1)} \right]^+
\end{split}
\end{equation}
where the superscript $(t)$ denotes the iteration index, $\varepsilon$ and $\tau$ are stepsizes, and $[\cdot]^+=\max\{0,\cdot\}$.

\subsection{Cram\'er-Rao Lower Bound}
This subsection briefly derives the CRLBs as benchmark for performance evaluation. For an $p\times 1$ estimator $\hat{\boldsymbol \theta}$, the Cram\'er-Rao theorem \cite{Kay} states
\begin{equation}\label{eq:CR}
{\mathop{\rm var}} \left( {{\theta}_m} \right) \ge {\left[ {{{\bf F}^{ - 1}}\left( \boldsymbol \theta \right)} \right]_{mm}}
\end{equation}
where ${{\bf F}\left( \boldsymbol \theta \right)}\in {\mathbb R}^{p\times p}$ represents the Fisher information matrix (FIM) and $\left[ {{{\bf F}^{ - 1}}\left( \boldsymbol \theta \right)} \right]_{mm}$ is the CRLB of the $m$th estimate. Assuming ${\bf y} \sim {\mathcal {CN}}(\boldsymbol\mu(\boldsymbol\theta),{\bf W}(\boldsymbol\theta))$, we express the $(m,n)$th entry of FIM as
\begin{equation}\label{eq:FIM}
\begin{split}
{\left[ {{\bf{F}}\left( {\boldsymbol{\theta }} \right)} \right]_{mn}} =& 2\Re \left( {\frac{{\partial {{\boldsymbol{\mu }}^{\rm{H}}}\left( {\boldsymbol\theta} \right)}}{{\partial {\theta _m}}}{{\bf{W}}^{ - 1}}\left( {\boldsymbol{\theta }} \right)\frac{{\partial {\boldsymbol{\mu }}\left( {\boldsymbol\theta} \right)}}{{\partial {\theta _n}}}} \right)\\
 &+ {\rm{tr}}\left( {{{\bf{W}}^{ - 1}}\left( {\boldsymbol{\theta }} \right)\frac{{\partial {\bf{W}}\left( {\boldsymbol{\theta }} \right)}}{{\partial {\theta _m}}}{{\bf{W}}^{ - 1}}\left( {\boldsymbol{\theta }} \right)\frac{{\partial {\bf{W}}\left( {\boldsymbol{\theta }} \right)}}{{\partial {\theta _n}}}} \right).
\end{split}
\end{equation}

To enable tractable analysis, we rewrite $\bf x$ as ${\bf z}=[\Re(h),\Im(h),\eta]^{\rm T}$. Suppose the probability density function of $\bf y$ is parameterized by real vector $\bf z$ with spatiotemporally uncorrelated noise, i.e., ${\bf w} \sim {\mathcal {CN}}\left( {{\bf 0},\sigma _w^2\bf I} \right)$, we have
\begin{equation}\label{eq:ysim}
{\bf y} \sim {\mathcal {CN}}({{\bf a}_1\Re(h)+j{\bf a}_1\Im(h)+{\bf a}_2\eta},{\sigma _w^2\bf I})
\end{equation}
where ${\bf a}_n$ denotes the $n$th column of $\bf A$.

Based on (\ref{eq:FIM}) and (\ref{eq:ysim}), the FIM of $\bf z$ is derived as
\begin{equation}\label{eq:F}
\begin{split}
{\bf F}\left( \bf z \right) =& \frac{2}{{\sigma _w^2}}\left[ {\begin{array}{*{20}{c}}
{{\bf a}_1^{\rm H}{{\bf a}_1}}&0&{{\bf a}_1^{\rm H}{{\bf a}_2}}\\
0&{{\bf a}_1^{\rm H}{{\bf a}_1}}&0\\
{{\bf a}_2^{\rm H}{{\bf a}_1}}&0&{{\bf a}_2^{\rm H}{{\bf a}_2}}
\end{array}} \right]\\
=& \frac{2}{{\sigma _w^2}}\left[ {\begin{array}{*{20}{c}}
{{\bf s}_{\rm p1}^{\rm H}{{\bf s}_{\rm p1}} + {\bf s}_{\rm p2}^{\rm H}{{\bf s}_{\rm p2}}}&0&{{\bf s}_{\rm p2}^{\rm H}{{\bf s}_{\rm p2}}}\\
0&{{\bf s}_{\rm p1}^{\rm H}{{\bf s}_{\rm p1}} + {\bf s}_{\rm p2}^{\rm H}{{\bf s}_{\rm p2}}}&0\\
{{\bf s}_{\rm p2}^{\rm H}{{\bf s}_{\rm p2}}}&0&{{\bf s}_{\rm p2}^{\rm H}{{\bf s}_{\rm p2}}}
\end{array}} \right]
\end{split}
\end{equation}
yielding
\begin{equation}\label{eq:Finv}
{{\bf F}^{ - 1}}\left(\bf z \right) = \frac{{\sigma _w^2}}{2}\left[ {\begin{array}{*{20}{c}}
{\frac{1}{{{\bf s}_{\rm p1}^{\rm H}{{\bf s}_{\rm p1}}}}}&0&{\frac{1}{{{\bf s}_{\rm p1}^{\rm H}{{\bf s}_{\rm p1}}}}}\\
0&{\frac{1}{{{\bf s}_{\rm p1}^{\rm H}{{\bf s}_{\rm p1}} + {\bf s}_{\rm p2}^{\rm H}{{\bf s}_{\rm p2}}}}}&0\\
{\frac{1}{{{\bf s}_{\rm p1}^{\rm H}{{\bf s}_{\rm p1}}}}}&0&{\frac{1}{{{\bf s}_{\rm p1}^{\rm H}{{\bf s}_{\rm p1}}}} + \frac{1}{{{\bf s}_{\rm p2}^{\rm H}{{\bf s}_{\rm p2}}}}}
\end{array}} \right].
\end{equation}

From (\ref{eq:Finv}), it follows that the CRLBs of $\Re{(h)}$, $\Im{(h)}$ and $\eta$ are equal to
\begin{equation}\label{eq:CRLBReh}
{\mathop{\rm var}} ({\Re\hat(h)}) \ge {\left[ {{{\bf F}^{ - 1}}\left(\bf z \right)} \right]_{11}} =  \frac{\sigma _w^2}{2{\bf s}_{\rm p1}^{\rm H}{{\bf s}_{\rm p1}}},
\end{equation}
\begin{equation}\label{eq:CRLBImh}
{\mathop{\rm var}} ({\Im\hat(h)}) \ge {\left[ {{{\bf F}^{ - 1}}\left(\bf z \right)} \right]_{22}} =  \frac{\sigma _w^2}{2\left({\bf s}_{\rm p1}^{\rm H}{{\bf s}_{\rm p1}}+{\bf s}_{\rm p2}^{\rm H}{{\bf s}_{\rm p2}}\right)},
\end{equation}
\begin{equation}\label{eq:CRLBeta}
{\mathop{\rm var}} \left( {\hat \eta } \right) \ge {\left[ {{{\bf F}^{ - 1}}\left( \bf z \right)} \right]_{33}} = \frac{\sigma _w^2}{2{\bf s}_{\rm p1}^{\rm H}{\bf s}_{\rm p1}}+\frac{\sigma _w^2}{2{\bf s}_{\rm p2}^{\rm H}{\bf s}_{\rm p2}}.
\end{equation}

\section{Numerical Results}
In this section, we conduct simulation to evaluate the mean squared error (MSE) performance for both designed estimation scheme (DES) and LS. Parameters used in the simulation are summarized as follows: variance of channel distributions $\sigma_h^2=\frac{1}{64}$, $\sigma_f^2=\frac{1}{25}$, $\sigma_g^2=\frac{1}{9}$; length of pilot $K_{\rm p}=2$, $K_1=1$; number of reflecting elements $N=32$; maximum number of iterations $t_{\max}=50$, maximum tolerance for convergence $\varpi=10^{-3}$. The results are averaged over $10^4$ independent Monte Carlo simulations.

\begin{figure}[t]
\centering
\includegraphics[scale=0.6]{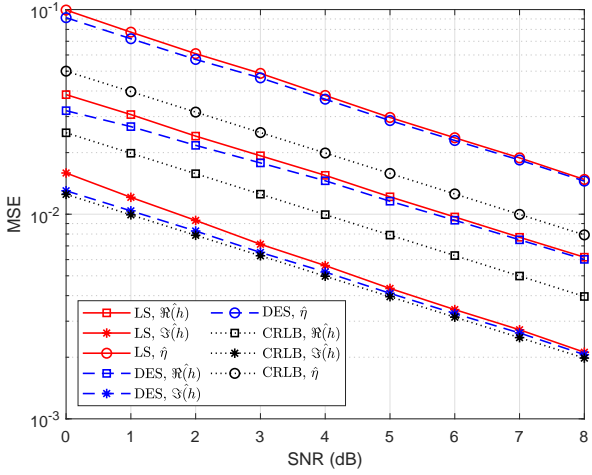}
\caption{MSEs versus SNR obtained from DES, LS and CRLB. }
\label{fig:fig3}
\end{figure}

In Fig. \ref{fig:fig3}, the MSE curves versus signal-to-noise ratio (SNR) is depicted for DES. It can be seen from Fig. \ref{fig:fig3} that MSE curves decline with the increase of SNR, and the gap between each MSE curve and corresponding CRLB remains unchanged in logarithmic ordinates, indicating that MSEs are prone to converge to CRLB as SNR goes adequately high. Comparing DES and LS, it is clear that the DES estimate (red solid line) perform better than LS estimate (blue dash line), which illustrates that estimation accuracy could be improved by taking the positive nature and superiority of assistant channel over direct channel into consideration.

We also quantize the estimation accuracy gains of DES over LS in Table \ref{Tab1}. We can see that $\hat\Re{(h)}$, $\hat\Im{(h)}$ and $\hat\eta$ peak at point SNR $=0$ dB with corresponding gains of $16.53\%$, $18.12\%$ and $8.24\%$, while the estimation accuracy improvement are less significant, around $2\%$, at point SNR$=8$ dB.

\begin{table}[t]
\centering
\caption{Estimation accuracy gains of DES over LS}\label{Tab1}
\begin{tabular}{c c c c c c c}
\hline
\hline
\multicolumn{2}{c}{SNR(dB)}  & 0              & 2 & 4 & 6 & 8  \\ \hline
\multirow{3}{*}{Gains (\%)}  & $\hat\Re(h)$   & 16.53 & 9.87 & 5.92 & 3.61 & 2.41  \\
                             & $\hat\Im(h)$   & 18.12 & 11.22 & 6.99 & 4.39 & 2.67 \\
                             & $\hat\eta$     & 8.24 & 6.29 & 4.11 & 3.03 & 1.78 \\ \hline\hline
\end{tabular}
\end{table}

\begin{figure}[t]
\centering
\includegraphics[scale=0.6]{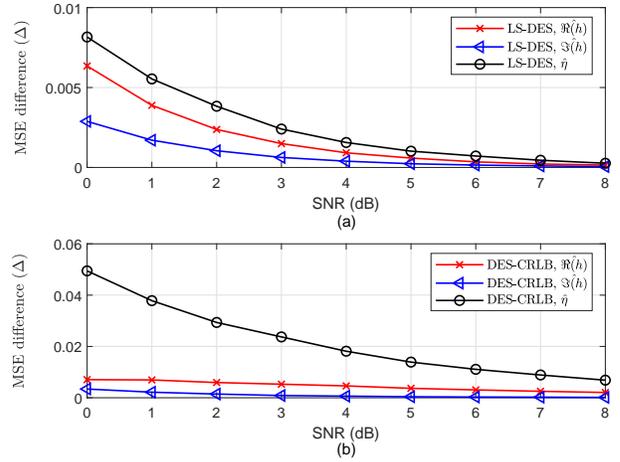}
\caption{MSE differences versus SNR obtained from (a) DES and LS; (b) DES and derived CRLB.}
\label{fig:fig4}
\end{figure}

Fig. \ref{fig:fig4} shows MSE difference curves versus SNR obtained from DES, LS and derived CRLB. The differences are calculated between various schemes, e.g., $\Delta_{\rm LS-DES} = {\rm MSE}_{\rm LS} - {\rm MSE}_{\rm DES}$. Fig.\ref{fig:fig4} (a) shows the persistent performance gaps between LS and DES as SNR varies from 0 to 8 dB, stating the superiority of DES over LS. With regard to Fig.\ref{fig:fig4} (b), the MSE differences between DES and CRLB are consistently positive, which demonstrates that CRLB curves lie below MSE curves of DES and further verifies the validity of our theoretical derivations in Section III-B.

\begin{figure}[t]
\centering
\includegraphics[scale=0.6]{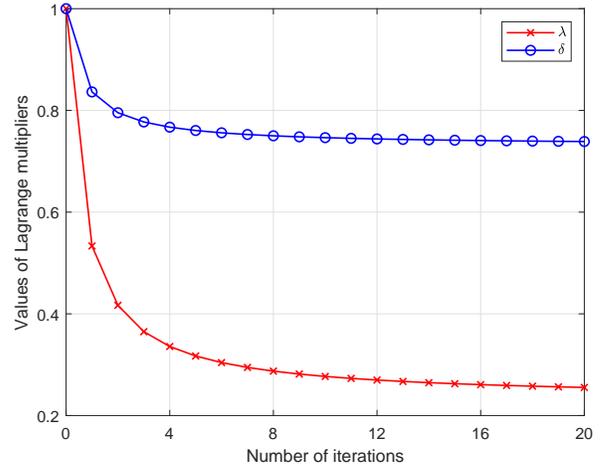}
\caption{Number of iterations versus values of Lagrange multipliers $\lambda$ and $\delta$.}
\label{fig:fig5}
\end{figure}

Fig. \ref{fig:fig5} presents the values of the Lagrange multiplier $\lambda $ and $\delta$ versus number of iterations. It can be observed from Fig. \ref{fig:fig5} that the Lagrange multipliers remain constant quickly after around 10 and 16 iterations, respectively, which indicates that DES offers quick convergence to globally optimal estimates.

\section{Conclusion}
In this letter, we have shown that CE for wireless communication systems assisted by LIS is a constrained estimation error minimization problem due to the features of assistant channel, which is different from traditional channel estimation problems. A Lagrange multiplier and dual ascent-based estimation scheme has been designed to solve the problem and obtain the closed-form expression for the combined channel parameters iteratively. In addition, the corresponding CRLBs have been deduced for performance evaluation. Numerical results have demonstrated that estimation accuracy of DES is, at most, 18$\%$ higher than that of LS in low SNR regime.

\end{document}